\documentclass[10pt,conference]{IEEEtran}

\usepackage{booktabs}
\usepackage{multirow}
\usepackage{tabularx}
\usepackage{amsmath}
\usepackage{amssymb}
\usepackage{url}
\usepackage{xspace}
\usepackage{tikz}
\usetikzlibrary{arrows.meta,positioning,patterns}
\usepackage{pgfplots}
\pgfplotsset{compat=1.18}
\usepackage[ruled,vlined,linesnumbered]{algorithm2e}
\usepackage{xcolor}
\usepackage[most]{tcolorbox}
\usepackage{listings}
\usepackage{pgf-pie}
\usetikzlibrary{fit,calc,shapes.symbols}
\usepackage[table]{xcolor}
\usepackage{array}
\usepackage{transparent}

\usepackage{pifont}

\newcommand{\cmark}{\textcolor{teal!70!black}{\ding{51}}}
\newcommand{\xmark}{\textcolor{red!70!black}{\ding{55}}}

\definecolor{bestbg}{HTML}{FFF3CD}

\newcommand{\Description}[1]{}

\definecolor{cBlue}{HTML}{4E79A7}
\definecolor{cOrange}{HTML}{F28E2B}
\definecolor{cGreen}{HTML}{59A14F}
\definecolor{cRed}{HTML}{E15759}
\definecolor{cTeal}{HTML}{76B7B2}
\definecolor{cPurple}{HTML}{B07AA1}
\definecolor{cYellow}{HTML}{EDC948}
\definecolor{cGray}{HTML}{BAB0AC}
\definecolor{cCyan}{HTML}{499894}

\lstdefinelanguage{json}{
  morestring=[b]",
  stringstyle=\color{blue!45!black},
  morecomment=[l]{//},
  commentstyle=\color{gray!70!black},
}
\lstdefinestyle{jsonsnippet}{
  language=json,
  basicstyle=\ttfamily\scriptsize,
  numbers=left,
  numberstyle=\tiny,
  stepnumber=1,
  numbersep=5pt,
  frame=single,
  breaklines=true,
  columns=fullflexible,
  keepspaces=true,
  showstringspaces=false
}

\newcommand{\benchmark}{R2Act\xspace}
\newcommand{\system}{Online Boutique\xspace}
\newcommand{\researchq}[1]{RQ#1}
\newtcolorbox{rqresultbox}{
  colback=gray!8,
  colframe=gray!45,
  boxrule=0.45pt,
  arc=2pt,
  left=4pt,
  right=4pt,
  top=3pt,
  bottom=3pt,
  before skip=6pt,
  after skip=6pt
}
\newcommand{\rqresult}[2]{\begin{rqresultbox}\noindent\textbf{Summary:} #2\end{rqresultbox}}

\begin{document}

\title{Can LLMs Really Recover Microservice Failures? A Recovery-Aware Evaluation of Diagnosis-to-Action Reasoning}

\author{
\IEEEauthorblockN{Jiaxing Qi\IEEEauthorrefmark{1}, Zhongzhi Luan\IEEEauthorrefmark{1},
Hongyu Zhang\IEEEauthorrefmark{2}, Shaohan Huang\IEEEauthorrefmark{3}\\
Carol Fung\IEEEauthorrefmark{4}, Yongxin Tong\IEEEauthorrefmark{1},
Hailong Yang\IEEEauthorrefmark{1}, and Depei Qian\IEEEauthorrefmark{1}}
\IEEEauthorblockA{\IEEEauthorrefmark{1}Beihang University\\
jiaxingqi@buaa.edu.cn, luan.zhongzhi@buaa.edu.cn, yxtong@buaa.edu.cn,
hailongyang@buaa.edu.cn, depeiq@buaa.edu.cn}
\IEEEauthorblockA{\IEEEauthorrefmark{2}Chongqing University\\
hyzhang@cqu.edu.cn}
\IEEEauthorblockA{\IEEEauthorrefmark{3}Microsoft Research Asia\\
shaohanh@microsoft.com}
\IEEEauthorblockA{\IEEEauthorrefmark{4}Concordia University\\
carol.fung@concordia.ca}
}

\maketitle

\begin{abstract}
Large language models (LLMs) are increasingly used to interpret operational evidence and assist incident response in cloud-native microservice systems. However, recovery-oriented use cases require more than identifying a root cause. After observing symptoms and diagnosing a fault, an operator or agent must translate the diagnosis into a concrete recovery action, apply it to an admissible target, and verify that service health has been restored. Existing RCA and log-analysis evaluations are well-suited to diagnosis, but they do not characterize this subsequent action decision. This paper presents R2Act, a recovery-action evaluation framework for post-diagnosis incident response. R2Act defines an incident schema, quality gate, action-space representation, recovery-validity metrics, offline evaluator, and live-replay protocol. We instantiate the framework as a benchmark dataset of 302 quality-audited Kubernetes incidents from \system. Each incident provides synchronized multi-modal observations, root-cause labels, an incident-specific action space, and annotated valid and invalid recovery plans. We evaluate heuristic, supervised, RCA-oriented, deep log, and LLM-based methods. The strongest RAG-based LLMs reach 91.4\%--99.7\% root-cause service accuracy, yet their recovery validity remains only 36.8\%--60.3\%. Even when both the root-cause service and fault type are correct, recovery-oriented methods still choose invalid actions for 39.5\%--62.0\% of correctly diagnosed incidents. In validity-gated live replay, 146 of 302 Qwen-RAG predictions are both offline-valid and recovered in live execution. Overall, this work reveals that many recovery failures arise not from missing diagnostic knowledge, but from the difficulty of translating diagnostic evidence into valid recovery actions and admissible targets. This work provides a reproducible, simplified starting point for research and evaluation.
\end{abstract}


\section{Introduction}
AI-assisted operations are advancing rapidly along several axes. LLM-based operational systems can summarize logs, retrieve similar incidents, localize likely root causes, and draft mitigation steps~\cite{sarda2024leveraging,zhong2026llm}. At the same time, cloud-native microservice systems have become harder to recover because independent deployment, elastic scaling, and dense service dependencies create more ways for failures to propagate~\cite{zhou2018benchmarking,liu2021microhecl,sun2025interpretable}. A user-visible incident may leave evidence scattered across logs, Kubernetes events, metrics, and resource state. These trends change incident response from a diagnosis-only task into an action-oriented decision problem: \textit{an automated method must not only explain what failed, but also decide which recovery action should be taken and where that action should be applied.}
\begin{figure}[t]
    \centering
    \includegraphics[width=\linewidth]{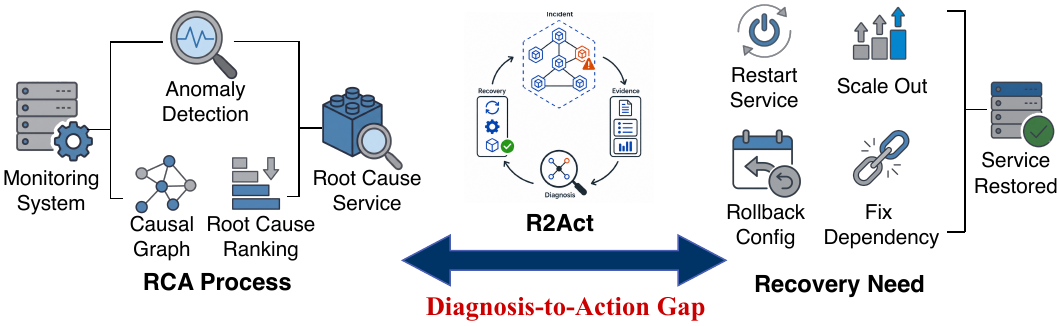}
    \caption{R2Act adds a recovery-action evaluation layer after diagnosis, checking whether methods can select a valid operation and target under incident-specific constraints.}
    \Description{A diagram of R2Act. It adds a recovery-action evaluation layer after diagnosis and checks whether a method can select a valid operation and target.}
    \label{fig:scenario}
\end{figure}

Yet there have been relatively few efforts in this recovery process. Real microservice recovery begins with partial, noisy, and multi-modal operational evidence, while the required outcome is a concrete decision that may modify a live deployment (see Figure~\ref{fig:scenario} for a high-level schematic of the missing process). Reaching that decision requires a method to assess the available evidence, infer the affected service and fault mechanism, and choose an operation whose target is valid for the current incident. The central question is therefore not only whether a method can identify a root cause, but whether it can turn that diagnosis into a valid recovery action under incident-specific constraints.

Existing evaluations mostly isolate earlier stages or adjacent forms of this workflow. Log-analysis benchmarks measure parsing, anomaly detection, log understanding, or diagnosis~\cite{zhu2023loghub,jiang2024large,cui2025logeval}; RCA methods and benchmarks evaluate which component, service, or fault type caused the incident~\cite{wang2024mrca,pham2025rcaeval}; and LLM-based operational reasoning systems study retrieval, prompting, diagnostic data collection, or mitigation recommendation~\cite{xu2025logsage,barnes2026logsieve,xu2025openrca,wang2023rcagent,chen2024automatic}. Remediation and self-healing work moves closer to recovery execution from diagnosis reports or predefined policies~\cite{bucchiarone2022mape,sarda2024leveraging,zhang2025microremed}. These settings rarely combine synchronized incident evidence, typed recovery plans, incident-specific validity labels, live replay, and reusable evaluation artifacts. As a result, the post-diagnosis action decision remains under-specified.

This gap matters because diagnostic correctness does not imply recovery-action validity. A method may correctly identify that a service is affected by a DNS fault while still proposing a service restart instead of repairing the dependency-resolution path. An HTTP routing fault may require rolling back a traffic rule rather than restarting the diagnosed service. A memory-pressure incident may require changing a resource field rather than scaling an unrelated component. In such cases, the method is not simply missing a root-cause label; it fails to carry the implications of the diagnosis into a valid operation, target, and optional parameter choice.

To fill this gap, we introduce R2Act, a benchmark construction and evaluation framework for diagnosis-to-action reasoning in microservice failure recovery. R2Act defines an incident schema, a quality gate, an incident-specific action-space representation, recovery-validity metrics, an offline evaluator, and a live-replay protocol. We instantiate the framework as a benchmark dataset of 302 quality-audited Kubernetes incidents collected from \system. Each incident packages synchronized logs, events, metrics, Kubernetes state, root-cause labels, incident-specific recovery action spaces, and annotated valid recovery plans with invalid alternatives. R2Act is the evaluation framework, while the 302 incidents are the current benchmark dataset instance used in our study.

\begin{figure}[t]
  \centering
  \includegraphics[width=\columnwidth]{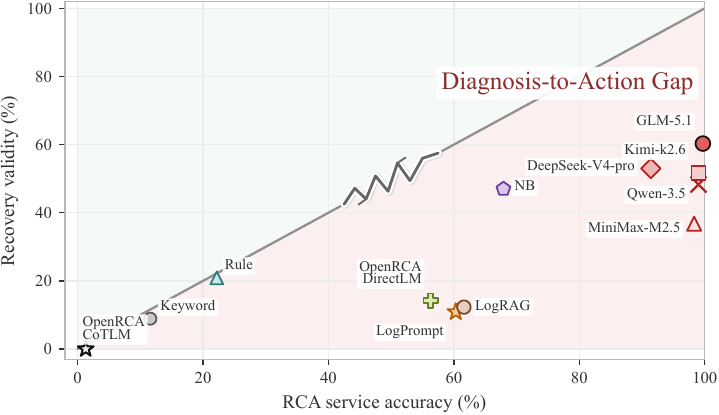}
  \caption{The plot compares diagnostic accuracy and recovery-action validity on the same incident population. Their divergence shows that recovery-oriented evaluation requires action-level metrics beyond diagnostic correctness.}
  \Description{Scatter plot comparing diagnostic accuracy and recovery-action validity across all compared methods, with an inset zooming into the low-score region.}
  \label{fig:diagnosis-recovery-gap}
\end{figure}

Figure~\ref{fig:diagnosis-recovery-gap} previews the main phenomenon measured by R2Act: the difference between diagnostic correctness and recovery-action validity when methods must select operation-target plans under incident-specific constraints. In our evaluations, heuristic baselines, supervised models, RCA-oriented techniques, deep log models, and LLM-based agents all show this diagnosis-to-action gap. The strongest RAG-based LLMs reach 91.4\%--99.7\% root-cause service accuracy, yet their recovery validity remains only 36.8\%--60.3\%. Even when both the root-cause service and fault type are correctly identified, recovery-oriented methods still choose invalid plans for 39.5\%--62.0\% of correctly diagnosed incidents. The errors concentrate in DNS, HTTP, and memory faults, where valid recovery depends on dependency, configuration, or resource semantics beyond the diagnosed service. In validity-gated live replay, 146 of 302 Qwen-RAG predictions are both offline-valid and recovered in live execution, showing that action-level validity bridges diagnostic evaluation and executable recovery.

This paper makes the following contributions:
\begin{itemize}
\item We formulate diagnosis-to-action recovery evaluation as a distinct problem for microservice incident response. The task asks whether a method can select a valid operation and admissible target under incident-specific constraints.
\item We present R2Act, a benchmark construction and evaluation framework with synchronized incident evidence, typed recovery action spaces, validity metrics, an offline evaluator, and a live-replay protocol. We instantiate it as a 302-incident Kubernetes benchmark covering six service roles and eight fault categories.
\item Extensive evaluation demonstrates that the diagnosis-to-action gap is a persistent failure mode in microservice recovery. Even after correct RCA, methods frequently select invalid recovery plans, with errors dominated by wrong operations and invalid plan structures. We further validate offline recovery decisions through live replay, which checks whether predicted actions actually restore service health.

\end{itemize}


\begin{table*}[t]
  \centering
  \scriptsize
  \setlength{\tabcolsep}{2.6pt}
  \renewcommand{\arraystretch}{1.0}
  \caption{Qualitative comparison with representative evaluation settings. The feature columns focus on recovery-oriented incident-response evaluation: synchronized incident evidence, root-cause labels, typed recovery plans, incident-specific validity labels, live recovery replay, and reusable evaluation artifacts.}
  \label{tab:related-comparison}

  \resizebox{0.98\textwidth}{!}{%
  \begin{tabular}{@{}
    >{\raggedright\arraybackslash}m{0.235\textwidth}
    >{\raggedright\arraybackslash}m{0.245\textwidth}
    >{\centering\arraybackslash}m{0.090\textwidth}
    >{\centering\arraybackslash}m{0.075\textwidth}
    >{\centering\arraybackslash}m{0.075\textwidth}
    >{\centering\arraybackslash}m{0.085\textwidth}
    >{\centering\arraybackslash}m{0.075\textwidth}
    >{\centering\arraybackslash}m{0.090\textwidth}
    @{}}
    \toprule
    \centering\textbf{Existing Works}
      & \centering\textbf{Primary Task}
      & \shortstack{\textbf{Multi-modal}\\\textbf{Incident}\\\textbf{Evidence}}
      & \shortstack{\textbf{Root-cause}\\\textbf{Labels}}
      & \shortstack{\textbf{Typed}\\\textbf{Recovery}\\\textbf{Plans}}
      & \shortstack{\textbf{Incident}\\\textbf{specific}\\\textbf{Validity}\\\textbf{Labels}}
      & \shortstack{\textbf{Live}\\\textbf{Recovery}\\\textbf{Replay}}
      & \shortstack{\textbf{Reusable}\\\textbf{Evaluation}\\\textbf{Artifact}} \\
    \midrule

    Microservice RCA benchmarks~\cite{pham2025rcaeval}
      & Root-cause localization from telemetry
      & \cmark
      & \cmark
      & \xmark
      & \xmark
      & \xmark
      & \cmark \\

    Log analysis and LLM reasoning~\cite{cui2025logeval,xu2025logsage,xu2025openrca}
      & Log understanding, anomaly detection, or root-cause analysis
      & \xmark
      & \cmark
      & \xmark
      & \xmark
      & \xmark
      & \xmark \\

    Self-healing and remediation methods~\cite{bucchiarone2022mape,sarda2024leveraging,zhang2025microremed}
      & Mitigation or remediation generation from diagnosis or policies
      & \xmark
      & \cmark
      & \cmark
      & \xmark
      & \cmark
      & \xmark \\

    Execution-oriented SE benchmarks~\cite{jimenez2024swe,bhuiyan2023secbench}
      & Executable validation of generated software artifacts
      & \xmark
      & \xmark
      & \xmark
      & \xmark
      & \xmark
      & \cmark \\

    \textbf{R2Act (Ours)}
      & \textbf{Diagnosis-to-action recovery evaluation}
      & \cmark
      & \cmark
      & \cmark
      & \cmark
      & \cmark
      & \cmark \\

    \bottomrule
  \end{tabular}%
  }
\end{table*}

\section{Related Work and Motivation}

\subsection{Related Work}

\textbf{Microservice RCA methods and evaluations.}
Microservice RCA has been studied with metric-based, causal-inference-based, log-based, and learning-based methods. These methods use causal graphs, random-walk ranking, hypothesis testing, metric anomaly analysis, or learned representations to localize faulty services or metrics. Recent studies and toolkits, including causal-inference RCA evaluations, PyRCA, and RCAEval, make these methods easier to compare on microservice telemetry~\cite{pham2024root,liu2023pyrca,pham2025rcaeval}. Their evaluation target is root-cause localization, which is the appropriate objective for diagnosis-stage RCA. R2Act adds a downstream action-level evaluation for recovery-oriented use cases.

\textbf{Log analysis and LLM-based operational reasoning.}
Log-analysis benchmarks evaluate parsing, anomaly detection, incident classification, and log-based diagnosis~\cite{zhu2023loghub,jiang2024large,cui2025logeval}. Recent LLM-based systems extend this direction with prompting, retrieval, and agent workflows for operational logs. LogSage and LogSieve study CI/CD failure analysis and log reduction, while LogPrompt, LogRAG, OpenRCA, and RCAFlow explore LLM-based log analysis or RCA~\cite{xu2025logsage,barnes2026logsieve,liu2024logprompt,zhang2024leveraging,xu2025openrca,gao2026rcaflow}. These studies show the value of LLMs for operational reasoning, but they mainly evaluate diagnosis, log understanding, or task-specific remediation support rather than incident-specific recovery-action validity.

\textbf{Self-healing and autonomic recovery methods.}
Recent work has started to connect cloud incident diagnosis with remediation~\cite{chen2025grace}. LLM-based systems such as mitigation-step recommendation, RCACopilot, and RCAgent support root-cause analysis, diagnostic data collection, or mitigation recommendation in production cloud settings~\cite{ahmed2023recommending,chen2024automatic,wang2023rcagent}. Other work moves closer to recovery execution: Sarda et al.~\cite{sarda2024leveraging} generate Ansible playbooks for microservice auto-remediation, GenKubeSec~\cite{malul2024genkubesec} studies Kubernetes misconfiguration remediation, and GALR~\cite{zhang2026galr} combines graph-based RCA with LLM-assisted recovery planning. MicroRemed~\cite{zhang2025microremed} is the closest benchmark to R2Act because it evaluates LLMs on executable microservice remediation from diagnosis reports. R2Act is complementary: it starts from synchronized multi-modal incident observations and evaluates whether methods can select valid recovery operations and admissible targets under incident-specific action spaces.
 
Prior remediation work often starts from diagnosis reports, mitigation suggestions, or predefined adaptation policies. R2Act starts from synchronized incident observations and asks whether methods can both diagnose the failure and select a valid operation and target for that specific incident. This setting connects diagnosis and recovery planning while keeping the post-diagnosis recovery-action decision explicit.

\textbf{Execution-oriented software engineering benchmarks.}
Execution-oriented benchmarks evaluate generated outputs through explicit checks rather than only textual similarity or classification accuracy. SWE-bench checks whether code changes resolve real GitHub issues, and SecBench provides executable security tests for server-side JavaScript~\cite{jimenez2024swe,bhuiyan2023secbench}. These benchmarks show that software engineering evaluation becomes more informative when outputs are validated against task-specific execution criteria.

\subsection{Motivation}
R2Act follows an execution-oriented benchmarking principle. A recovery plan should be evaluated not only by whether it identifies the relevant service, but also by whether its operation and target are valid for the specific incident. Table~\ref{tab:related-comparison} compares existing work along the dimensions required for recovery-oriented microservice evaluation. Prior benchmarks and systems have advanced diagnosis, log understanding, remediation recommendations, and executable software tasks. R2Act complements these efforts by introducing the action-level evaluation layer needed when incident-response methods are used to support or automate recovery. This comparison reveals two missing layers that remain under-specified in existing evaluations.

\textbf{Missing Layer 1: Post-diagnosis action-decision evaluation.}
Microservice recovery depends on heterogeneous evidence, including logs, events, metrics, and service state. Existing log or RCA benchmarks reasonably evaluate diagnosis-stage tasks such as log parsing~\cite{cui2025logeval}, anomaly detection~\cite{guo2024logformer}, log understanding~\cite{barnes2026logsieve}, and root-cause localization~\cite{pham2025rcaeval,gao2026rcaflow}. Recovery-oriented incident response introduces an additional downstream target: selecting a recovery operation and binding it to an admissible target under the current incident constraints.  

\textbf{Missing Layer 2: Incident-specific recovery constraints and admissible targets.}
RCA provides important evidence for recovery, but service and fault-type labels do not fully specify the recovery decision. A method can identify the correct service and fault type while still selecting an invalid operation or target. For example, an HTTP routing fault may require rolling back a traffic rule rather than restarting the diagnosed service. Such errors occur after diagnosis and determine whether the system actually recovers and whether an operator can safely approve the proposed change.

Given such landscape, R2Act connects multi-modal incident evidence, root-cause labels, incident-specific action spaces, valid and invalid recovery plans, and live replay evidence in one evaluation setting.

\begin{figure*}[t]
  \centering
  \includegraphics[width=\linewidth]{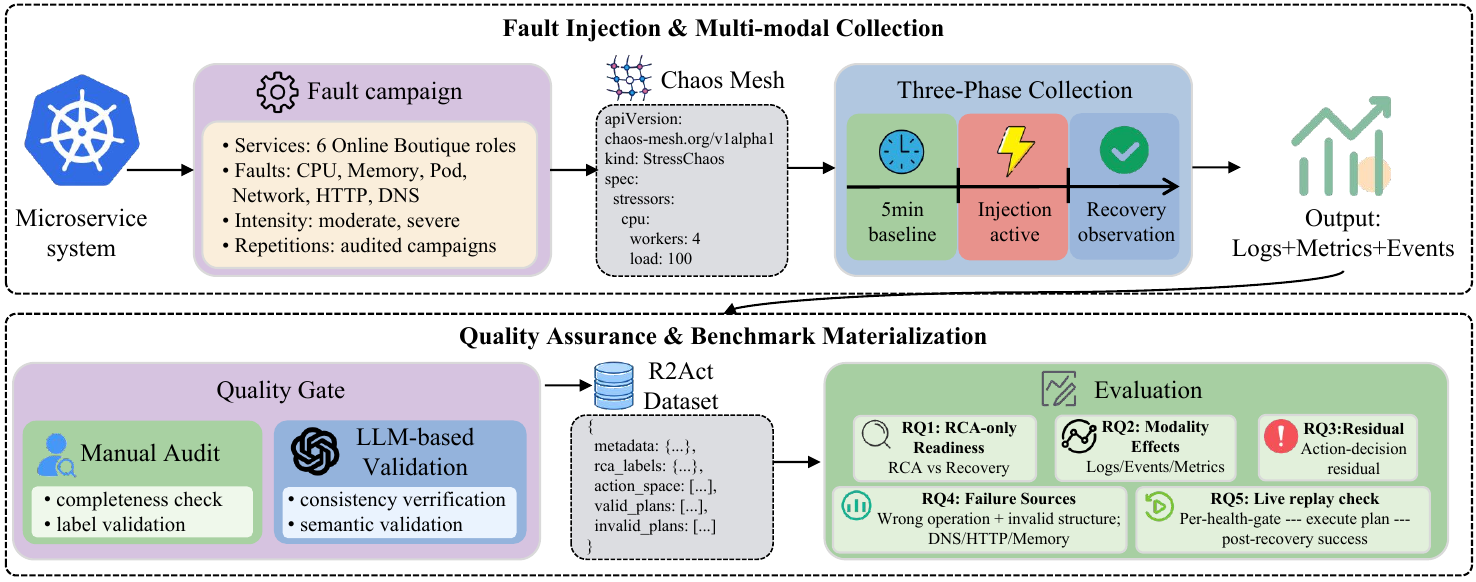}
  \caption{R2Act operationalizes an action-level evaluation layer for microservice incident response. It builds audited incidents from fault campaigns and synchronized multi-modal observations, materializes incident-specific action spaces and recovery plans, and evaluates post-diagnosis operation-target decisions through offline validity reports and live replay evidence.}
  \Description{The R2Act framework contains five modules: fault campaign design, live Kubernetes testbed, multi-modal observation plane, quality gate, and evaluation suite. The modules connect fault injection, synchronized observation collection, quality auditing, benchmark materialization, incident-specific action spaces, offline validity reports, and live replay evidence.}
  \label{fig:framework}
\end{figure*}

\section{Overview}

\benchmark is designed as an evaluation framework for recovery-oriented incident response. It preserves incident-time evidence, exposes the operational task, and makes the evaluation rule explicit. R2Act extends diagnosis-centered evaluation with an action-level recovery metric, where a prediction is valid only if its operation, target, and optional fields satisfy the incident-specific action space. Figure~\ref{fig:framework} presents the framework overview. R2Act defines fault campaigns over service roles, fault mechanisms, intensities, and repetitions, executes them in a live Kubernetes testbed, collects synchronized observations around each injected fault, and converts quality-audited cases into benchmark records and evaluation reports.

\subsection{Benchmark Design}

\benchmark uses \system~\cite{onlineboutique}, deployed on a Kubernetes cluster~\cite{kubernetes}.
The main benchmark covers six services that represent different roles in a microservice application: the entry-point \texttt{frontend}, the orchestration service \texttt{checkoutservice}, the recommendation and computation service \texttt{recommendationservice}, the state-related \texttt{cartservice}, the catalog read service \texttt{productcatalogservice}, and the transaction-path \texttt{paymentservice}. This coverage avoids concentrating the benchmark on a small set of structurally similar services.

In addition, \benchmark covers eight fault categories, organized into three broader mechanism groups. (1) Kubernetes lifecycle and resource faults include pod failure, CPU saturation, and memory pressure. (2) Network and application-layer faults include network delay, HTTP error, and HTTP abort. (3) Dependency-resolution faults include DNS fault and DNS randomization. This grouping is used only to explain the fault mechanisms; the evaluation still treats the eight fault categories as distinct labels. Most generated cases use moderate and severe intensity levels. Fault campaigns that did not pass the quality audit, including unstable I/O-pressure cases and incomplete network-delay runs, are retained in the audit trail but excluded from the main results.

Formally, each incident is represented as a record:
\begin{equation}
x_i=(O_i,M_i,Y_i,A_i),
\label{eq:record}
\end{equation}
where $O_i$ denotes operational observations, $M_i$ records campaign metadata, $Y_i$ contains RCA and recovery labels, and $A_i$ defines the allowed recovery-action space.
The observation component is multi-modal:
\begin{equation}
O_i=\{O_i^{log}, O_i^{event}, O_i^{metric}, O_i^{state}, O_i^{chaos}\}.
\label{eq:observation}
\end{equation}

These terms correspond to service logs, Kubernetes events, Prometheus metrics, pod/deployment state, and Chaos Mesh status.
This notation makes explicit that log-only, log-event, and full multi-modal settings are different views of the same incident record rather than different datasets.

\subsection{Construction and Quality Audit}

For each planned incident, the collector first checks that all deployments are ready, that the frontend health check succeeds, that no leftover Chaos Mesh object remains, and that Prometheus is reachable.
It then records a pre-fault snapshot, applies the corresponding Chaos Mesh fault object, waits for the injection window, records a during-fault snapshot, removes the fault object, verifies recovery, and records a post-fault snapshot.
Each run receives a stable incident identifier encoding the system, target service, fault mechanism, intensity, and repetition id.
The raw observation directory is therefore traceable back to the campaign manifest and the generated benchmark JSON.
Each incident is collected in three phases: pre-fault, during-fault, and post-fault.
For each phase, \benchmark records service logs from Kubernetes pods, Kubernetes events and pod/deployment states, Prometheus metrics~\cite{prometheus} with resource and service-level indicators, and Chaos Mesh status~\cite{chaosmesh} with the injected target and fault state.
The synchronized view supports log-only methods as well as methods that use events and metrics.

Quality control is applied before any incident enters the main dataset.
An incident is admitted only when all three observation phases exist, logs are present, metrics and Chaos Mesh status are present, Kubernetes state artifacts are present, the annotation JSON is valid, and the benchmark JSON can be built.
We model this decision as a quality gate:
\begin{equation}
q_i=\prod_{m\in\mathcal{M}}\mathbb{1}[O_i^m \neq \emptyset]\cdot \mathbb{1}[Y_i\ \mathrm{valid}]\cdot \mathbb{1}[A_i\ \mathrm{valid}],
\label{eq:quality-gate}
\end{equation}
where $\mathcal{M}$ is the set of required modalities.
Only incidents with $q_i=1$ enter the main benchmark.
The audit assigns every incident to one of four groups:
quality-audited main incidents, warning incidents, failed incidents, or legacy pilot incidents.
Warning and failed incidents are preserved for transparency, but they are excluded from the main evaluation tables.
This policy avoids mixing incomplete or unstable observations into the primary benchmark while retaining enough metadata to diagnose collection failures.

\subsection{Benchmark Records and Recovery Semantics}

After quality auditing, \benchmark represents each incident as a standardized benchmark instance containing observations, metadata, labels, action-space constraints, valid recovery plans, and invalid recovery plans.
Each incident is annotated with the root-cause service, root-cause fault type, fault intensity, fault mechanism, and recovery labels.
Recovery plans in R2Act are represented as typed operation-target decisions.
Each plan contains an operation type, a primary target service, and optional fields required by the operation, such as a dependency endpoint, configuration scope, route rule, or resource limit.
The incident-specific action space enumerates the operations and targets that are admissible under the injected fault and the current deployment state.
The valid recovery plan for each incident is constructed from the fault injection specification, the affected service-dependency or Kubernetes object, and the expected post-recovery system state.

Table~\ref{tab:recovery-actions} lists the concrete recovery actions represented in the current benchmark dataset.
For example, a DNS fault injected on the dependency path of cartservice is labeled with a dependency-resolution action rather than a generic service restart, because restarting the diagnosed service does not repair the injected DNS rule.
HTTP route faults are labeled with configuration-level recovery actions, and memory-pressure faults are labeled with resource-limit actions when the deployment constraint requires it.
Invalid plans are constructed to cover common recovery failure modes, including wrong operation, wrong target, invalid dependency or configuration scope, invalid plan structure, and no-op actions.
They are not random negative samples; they test whether a method distinguishes service localization from valid recovery semantics.
The evaluation distinguishes target-service correctness, action-type correctness, exact plan match, recovery validity, and no-op or irrelevant actions.
For a predicted plan $\hat{p}_i$, recovery validity is defined over the incident-specific action space, valid plan set $P_i^+$, and invalid plan set $P_i^-$:
\begin{equation}
\mathrm{valid}(\hat{p}_i)=\mathbb{1}[\hat{p}_i\in P_i^+ \land \hat{p}_i\notin P_i^- \land \hat{p}_i\in A_i].
\label{eq:recovery-validity}
\end{equation}

\begin{table}[t]
  \centering
  \caption{Recovery actions represented in R2Act.}
  \label{tab:recovery-actions}
  \scriptsize
  \setlength{\tabcolsep}{2.8pt}
  \resizebox{\linewidth}{!}{
  \begin{tabular}{lll}
    \toprule
    \textbf{Action} & \textbf{Typical faults} & \textbf{Required target fields} \\
    \midrule
    Restart service & Pod or delay faults & Target service \\
    Scale out & CPU saturation & Target service and replicas \\
    Roll back deployment & Deployment rollback cases & Target service \\
    Increase memory limit & Memory pressure & Target service and resource limit \\
    Roll back configuration & HTTP error or abort & Route or configuration scope \\
    Repair DNS/dependency & DNS fault or randomization & Service dependency endpoint \\
    No action & Invalid or no-op prediction & None \\
    \bottomrule
  \end{tabular}}
\end{table}

\begin{figure}[t]
  \centering
  \begin{tikzpicture}[
    block/.style={
      draw=blue!25,
      fill=blue!3,
      rounded corners=1pt,
      minimum width=0.98\columnwidth,
      align=left
    },
    block-red/.style={
      draw=red!30,
      fill=red!3,
      rounded corners=1pt,
      minimum width=0.98\columnwidth,
      align=left
    },
    block-green/.style={
      draw=green!40,
      fill=green!3,
      rounded corners=1pt,
      minimum width=0.98\columnwidth,
      align=left
    },
    header/.style={
      fill=blue!65!black,
      text=white,
      font=\bfseries\footnotesize,
      inner xsep=4pt,
      inner ysep=2pt,
      anchor=north west
    },
    header-red/.style={
      fill=red!65!black,
      text=white,
      font=\bfseries\footnotesize,
      inner xsep=4pt,
      inner ysep=2pt,
      anchor=north west
    },
    header-green/.style={
      fill=green!50!black,
      text=white,
      font=\bfseries\footnotesize,
      inner xsep=4pt,
      inner ysep=2pt,
      anchor=north west
    },
    body/.style={
      font=\ttfamily\scriptsize,
      inner xsep=4pt,
      inner ysep=3pt,
      anchor=north west,
      text width=0.93\columnwidth
    }
  ]
    \node[block-red,minimum height=1.3cm] (b1) at (0,0) {};
    \node[header-red] at (b1.north west) {Incident};
    \node[body] at ([yshift=-0.3cm]b1.north west) {%
      id="onlineboutique\_cartservice\_dns\_fault\_severe\_r02",\\
      service="cartservice", fault="dns\_fault", intensity="severe"};

    \node[block,minimum height=2.1cm,below=0.08cm of b1] (b2) {};
    \node[header] at (b2.north west) {Observation};
    \node[body] at ([yshift=-0.3cm]b2.north west) {%
      mechanism="chaos-mesh-dns-error", parameters="action=error, scope=service-domain, duration=120s";\\
      phases=[pre, during, post]; modalities=[logs, events];\\
      logs: 12 services, 43,847 entries, 12 pods; events: kubernetes state};

    \node[block-green,minimum height=1.6cm,below=0.08cm of b2] (b3) {};
    \node[header-green] at (b3.north west) {Data Structure};
    \node[body] at ([yshift=-0.3cm]b3.north west) {%
      \{metadata: \{batch, fault\_intensity, load\_profile, ...\},\\
      \hspace{0.5em}fault: \{fault\_type, target\_service, fault\_yaml\},\\
      \hspace{0.5em}input: \{logs: [\{phase, service, pod, level, message\}, ...], events: [...]\}\}};

    \node[block,minimum height=1.3cm,below=0.08cm of b3] (b4) {};
    \node[header] at (b4.north west) {Labels};
    \node[body] at ([yshift=-0.3cm]b4.north west) {%
      rca=("cartservice", "dns\_fault");\\
      valid action="fix\_dns\_resolution(cartservice)"; invalid action="restart\_service(frontend)"};

  \end{tikzpicture}
  \caption{Example standardized benchmark instance. The compact exhibit shows key fields of the severe DNS error,  while the released artifact retains the complete 43,847 log entries and structured annotations.}
  \Description{Single-column block-style exhibit showing key fields of a DNS fault benchmark record with data structure breakdown.}
  \label{fig:benchmark-instance}
\end{figure}
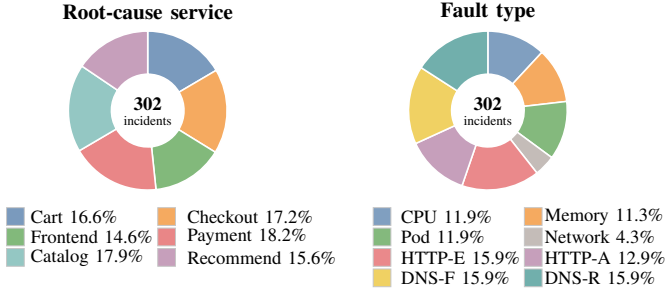
\begin{figure}[t]
  \centering
  \begin{tikzpicture}[
    font=\footnotesize,
    legendbox/.style={
      rectangle,
      draw=gray!45,
      line width=0.25pt,
      minimum width=0.16cm,
      minimum height=0.16cm
    },
    legendtext/.style={
      font=\scriptsize,
      anchor=west
    },
    title/.style={
      font=\bfseries\footnotesize,
      align=center
    }
  ]

    \def\outerR{1.05}
    \def\innerR{0.48}

    \newcommand{\donutsegment}[5]{%
      \path[
        fill=#5,
        draw=white,
        line width=0.6pt
      ]
        (#1,#2) -- ++(#3:\outerR)
        arc[start angle=#3,end angle=#4,radius=\outerR]
        -- ++(#4+180:\outerR-\innerR)
        arc[start angle=#4,end angle=#3,radius=\innerR]
        -- cycle;
    }

    \node[title] at (-2.25,1.55) {Root-cause service};

    \donutsegment{-2.25}{0.25}{90.0}{30.4}{cBlue!75}
    \donutsegment{-2.25}{0.25}{30.4}{-31.6}{cOrange!75}
    \donutsegment{-2.25}{0.25}{-31.6}{-84.0}{cGreen!75}
    \donutsegment{-2.25}{0.25}{-84.0}{-149.6}{cRed!72}
    \donutsegment{-2.25}{0.25}{-149.6}{-214.0}{cTeal!78}
    \donutsegment{-2.25}{0.25}{-214.0}{-270.0}{cPurple!72}

    \fill[white] (-2.25,0.25) circle (0.48);
    \node[font=\scriptsize\bfseries] at (-2.25,0.34) {302};
    \node[font=\tiny] at (-2.25,0.12) {incidents};

    \node[legendbox,fill=cBlue!75] at (-4.0,-1.15) {};
    \node[legendtext] at (-3.92,-1.15) {Cart 16.6\%};

    \node[legendbox,fill=cOrange!75] at (-2.0,-1.15) {};
    \node[legendtext] at (-1.85,-1.15) {Checkout 17.2\%};

    \node[legendbox,fill=cGreen!75] at (-4.0,-1.42) {};
    \node[legendtext] at (-3.92,-1.42) {Frontend 14.6\%};

    \node[legendbox,fill=cRed!72] at (-2.0,-1.42) {};
    \node[legendtext] at (-1.85,-1.42) {Payment 18.2\%};

    \node[legendbox,fill=cTeal!78] at (-4.0,-1.69) {};
    \node[legendtext] at (-3.92,-1.69) {Catalog 17.9\%};

    \node[legendbox,fill=cPurple!72] at (-2.0,-1.69) {};
    \node[legendtext] at (-1.85,-1.69) {Recommend 15.6\%};

    \node[title] at (2.25,1.55) {Fault type};

    \donutsegment{2.25}{0.25}{90.0}{47.1}{cBlue!72}
    \donutsegment{2.25}{0.25}{47.1}{6.6}{cOrange!75}
    \donutsegment{2.25}{0.25}{6.6}{-36.3}{cGreen!74}
    \donutsegment{2.25}{0.25}{-36.3}{-51.8}{cGray!85}
    \donutsegment{2.25}{0.25}{-51.8}{-109.0}{cRed!70}
    \donutsegment{2.25}{0.25}{-109.0}{-155.5}{cPurple!72}
    \donutsegment{2.25}{0.25}{-155.5}{-212.7}{cYellow!85}
    \donutsegment{2.25}{0.25}{-212.7}{-270.0}{cCyan!78}

    \fill[white] (2.25,0.25) circle (0.48);
    \node[font=\scriptsize\bfseries] at (2.25,0.34) {302};
    \node[font=\tiny] at (2.25,0.12) {incidents};

    \node[legendbox,fill=cBlue!72] at (0.85,-1.15) {};
    \node[legendtext] at (0.98,-1.15) {CPU 11.9\%};

    \node[legendbox,fill=cOrange!75] at (2.85,-1.15) {};
    \node[legendtext] at (2.88,-1.15) {Memory 11.3\%};

    \node[legendbox,fill=cGreen!74] at (0.85,-1.42) {};
    \node[legendtext] at (0.98,-1.42) {Pod 11.9\%};

    \node[legendbox,fill=cGray!85] at (2.85,-1.42) {};
    \node[legendtext] at (2.88,-1.42) {Network 4.3\%};

    \node[legendbox,fill=cRed!70] at (0.85,-1.69) {};
    \node[legendtext] at (0.98,-1.69) {HTTP-E 15.9\%};

    \node[legendbox,fill=cPurple!72] at (2.85,-1.69) {};
    \node[legendtext] at (2.88,-1.69) {HTTP-A 12.9\%};

    \node[legendbox,fill=cYellow!85] at (0.85,-1.96) {};
    \node[legendtext] at (0.98,-1.96) {DNS-F 15.9\%};

    \node[legendbox,fill=cCyan!78] at (2.85,-1.96) {};
    \node[legendtext] at (2.88,-1.96) {DNS-R 15.9\%};

  \end{tikzpicture}
  \caption{Proportional composition of the 302 quality-audited incidents by root-cause service and fault type. The donut charts show category proportions in the main benchmark. HTTP-E, HTTP-A, DNS-F, and DNS-R denote HTTP error, HTTP abort, DNS fault, and DNS random, respectively.}  
  \Description{Two donut charts showing the proportional distribution of incidents by root-cause service and fault type.}
  \label{fig:dataset-distribution}
\end{figure}
Here, $P_i^+$ is defined as a set rather than a single textual label. This set-valued allows one incident to admit multiple semantically equivalent recovery plans, and the evaluator accepts a prediction if it matches any member of $P_i^+$. Additional equivalent plans can be incorporated when they satisfy the same recovery-state constraint and are validated under the same action
schema.

All baselines and LLM-based methods are evaluated on the same benchmark records, which separates the cost of data collection from the evaluation of new methods. Once an incident has passed the quality audit and been materialized as a record, it can be reused without re-running fault injection. New recovery methods can therefore be compared under identical evidence and labels, while new campaigns can be appended through the same manifest-driven fault specification and quality-audit pipeline.

\begin{figure*}[t]
    \centering
    \includegraphics[width=1.0\textwidth]{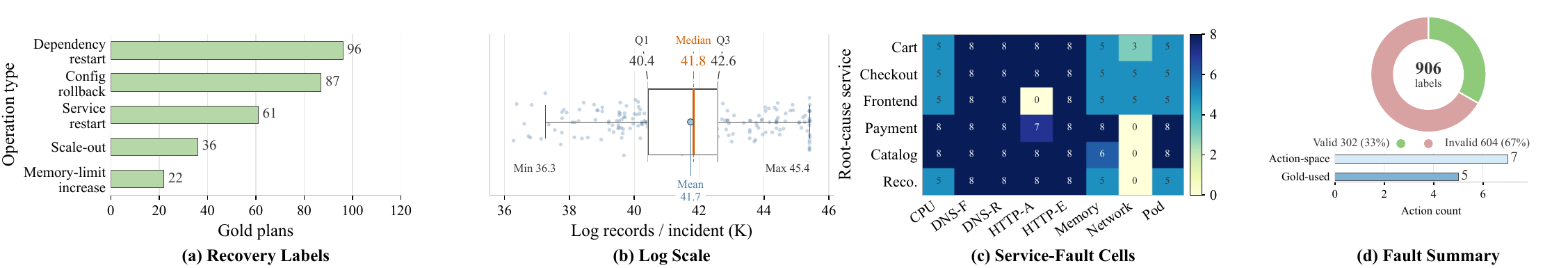}
    \caption{Characterization of the benchmark dataset. (a) Distribution of gold recovery plans across operation types. (b) Log-scale distribution of root-cause services and fault types. (c) Heatmap of service-fault cells, where color intensity indicates the number of incidents in each cell. (d) Donut chart showing the distribution of fault types.}
    \Description{Characterization of the benchmark dataset. (a) Distribution of gold recovery plans across operation types. (b) Log-scale distribution of root-cause services and fault types. (c) Heatmap of service-fault cells, where color intensity indicates the number of incidents in each cell. (d) Donut chart showing the distribution of fault types.}
    \label{fig:dataset_characterization}
\end{figure*}
\begin{table}[t]
  \centering
  \caption{Dataset summary and quality audit.}
  \label{tab:dataset-quality}
  \small
  \resizebox{\linewidth}{!}{
  \begin{tabular}{lrl}
    \toprule
    \textbf{Statistic} & \textbf{Value} & \textbf{Meaning} \\
    \midrule
    Main incidents & 302 & Used in all reported results \\
    Root-cause services & 6 & Distinct service roles \\
    Fault categories & 8 & Distinct RCA-type labels \\
    Fault mechanisms & 8 & Distinct injection mechanisms \\
    Service-fault cells & 44 & Non-empty service--fault pairs \\
    Cell size & 3--8 & Incidents per service--fault cell \\
    Moderate incidents & 129 & Medium fault intensity \\
    Severe incidents & 173 & High fault intensity \\
    \midrule
    Log records & 12.6M & Total main-set log volume \\
    Logs per incident & 41{,}736 & Average observation scale \\
    Metrics-present samples & 302 & Prometheus data available \\
    Chaos-status samples & 302 & Injection status available \\
    \midrule
    Candidate operations & 7 & Shared recovery action space \\
    Gold operation types & 5 & Operations used by valid plans \\
    Invalid examples & 604 & Two invalid plans per incident \\
    \midrule
    Non-main audit-trail incidents & 71 & Warning, failed, and pilot runs \\
    \bottomrule
  \end{tabular}}
\end{table}

\section{Dataset Characterization}
\label{sec:dataset}

\subsection{Dataset Scope and Quality Audit}

Table~\ref{tab:dataset-quality} summarizes the main dataset and the audit trail retained during construction.
The released benchmark contains 302 quality-audited incidents.
Each main incident includes logs, Kubernetes state, metrics, Chaos Mesh status, annotation JSON, and a standardized benchmark instance.
The quality gate excludes incomplete, unstable, or pilot-stage incidents from the primary evaluation.
The 71 non-main audit-trail incidents consist of warning, failed, and legacy pilot runs.
They are preserved for transparency but are not used in any reported evaluation result.
This separation is important for reproducibility: the main benchmark defines the evaluation population, while the audit trail documents why certain collection attempts were not admitted.

Figure~\ref{fig:benchmark-instance} shows a compact example of one standardized benchmark instance.
The figure reports only the key fields needed to understand the released schema.
The complete artifact keeps the full observation files, including the raw log window and structured annotation files.
This example clarifies that \benchmark is not a collection of isolated log snippets.
Each record connects a concrete fault to multi-modal observations, RCA labels, and recovery semantics.

\subsection{Service and Fault-Type Composition}

Figure~\ref{fig:dataset-distribution} shows the proportional composition of the main dataset. The service composition is close to balanced: each of the six root-cause services contributes between 44 and 55 incidents. These services cover different system roles, including the entry point, orchestration service, recommendation/computation service, state-related service, catalog-read service, and transaction-path service. The fault composition covers eight categories across lifecycle, resource, network, application-layer, and dependency-resolution mechanisms. The smaller network-delay slice reflects the quality audit rather than a design omission, because unstable network-delay cases were excluded from the main benchmark.
Together, Table~\ref{tab:dataset-quality} and Figure~\ref{fig:dataset-distribution} show that the dataset is not a repetition of a single service or a single failure mechanism.
The quality gate keeps incomplete incidents out of the primary evaluation, while the retained composition still covers different service roles and fault mechanisms.

\subsection{Recovery Labels, Observation Scale, and Cell Coverage}

Figure~\ref{fig:dataset_characterization} gives three additional views of the dataset.
\textbf{Figure~\ref{fig:dataset_characterization}(a)}: the recovery-label separates the shared action-space from the gold recovery labels.
The action-space contains seven candidate operation types, while the 302 gold valid recovery plans use five of them; the remaining operations remain available candidates but are not gold actions for the main fault incidents.
\textbf{Figure~\ref{fig:dataset_characterization}(b)}: each incident contains a substantial observation window rather than a short synthetic message.
The 302 incidents contain 12.6M log records, with an average of 41{,}736.47 records per incident.
\textbf{Figure~\ref{fig:dataset_characterization}(c)}: The service-fault heatmap characterizes the diversity of the dataset beyond the marginal service and fault distributions. It shows that the benchmark covers 44 distinct service-fault cells across six services and eight fault categories, with 3 to 8 incidents in each cell. This diversity is important for evaluating methods across different failure modes and for interpreting per-fault performance differences.


\rqresult{}{R2Act provides 302 quality-audited incidents over six service roles and eight fault categories. Each main incident contains multi-modal observations, structured RCA labels, recovery labels, and a shared action-space, enabling us to compare diagnosis and recovery behavior over the same incident population.}

\section{Evaluation Design}
\label{sec:experimental-design}
The experiments evaluate methods through the recovery-action evaluation layer provided by R2Act; they do not evaluate R2Act as a recovery method. Our goal is to examine whether recovery-oriented incident response can be reliably assessed beyond RCA labels. We therefore organize the evaluation in a progressive way. We first compare diagnostic correctness with recovery-action validity on the same incidents to test whether they measure different capabilities. We then study whether adding richer incident evidence can close this gap. Next, we condition on correctly diagnosed incidents to isolate the remaining post-diagnosis action-decision errors. We further analyze where these errors occur across fault types and error categories. Finally, we check whether offline recovery validity is consistent with live replay behavior. We investigate the following research questions:
\begin{itemize}
\item \textbf{RQ1}: How different are diagnostic correctness and recovery-action validity across methods?
\item \textbf{RQ2}: How does multi-modal incident evidence affect diagnosis and recovery-action validity?
\item \textbf{RQ3}: What post-diagnosis action-decision errors remain after correct diagnosis?
\item \textbf{RQ4}: Which fault types and error categories make recovery-action selection difficult?
\item \textbf{RQ5}: How well does offline action validity align with live replay outcomes?
\end{itemize}

\textbf{Baselines.}
(1) \textit{Recovery-oriented methods} include log-based LLM methods, LogPrompt~\cite{liu2024logprompt} and LogRAG~\cite{zhang2024leveraging}; OpenRCA variants~\cite{xu2025openrca}; and five LLM backbones evaluated under zero-shot, few-shot, and retrieval-augmented protocols. (2) \textit{RCA-only methods} include Keyword, Rule, supervised Naive Bayes, PyRCA methods~\cite{liu2023pyrca} (Epsilon Diagnosis, RCD, and Random Walk), RCAEval methods~\cite{pham2025rcaeval} (BARO and CIRCA), and deep log diagnosis models LogFormer~\cite{guo2024logformer} and OneLog~\cite{hashemi2024onelog}.  No-training methods are evaluated once on all incidents; learning-based methods are evaluated under a 5-fold cross-validation. 

\textbf{RCA-to-Action Mapper.}
For methods that output only RCA results, we use a RCA-to-action mapper as an evaluation bridge. The mapper converts a predicted root-cause service $\hat{s}_i$ and fault type $\hat{t}_i$ into a structured recovery plan $\hat{p}_i=g(\hat{s}_i,\hat{t}_i)$ within the incident-specific action space. It follows a fixed fault-type policy: pod failure and network delay trigger service restart, CPU saturation triggers scale-out, memory pressure triggers memory-limit increase, HTTP faults trigger configuration rollback, and DNS faults trigger dependency restart. Required route, configuration, or dependency fields are filled from admissible targets in the incident action space.

The mapper is not intended to represent an optimal recovery policy. It is used to separate two questions: how far coarse service/type diagnosis can be carried by a transparent rule, and where incident-specific action semantics are still required. We apply the mapper only to RCA-only methods, including keyword matching, rule-based variants, supervised Naive Bayes, PyRCA, RCAEval, LogFormer, and OneLog. Their action scores therefore measure the downstream recoverability of RCA outputs under this fixed policy, rather than the methods' own recovery-generation ability. Recovery-oriented methods, such as LogPrompt, LogRAG, OpenRCA, and LLM backbones, are evaluated using their generated recovery plans directly. Gold RCA + Mapper serves as a diagnostic control for the limits of coarse service/type-to-action mapping, not as a recovery upper bound.

\textbf{Evaluation Metrics.}
For diagnosis, \textbf{RCA-Service accuracy} measures the fraction of incidents whose predicted root-cause service matches the label, and \textbf{RCA-Type accuracy} measures the fraction whose predicted fault type matches the label. For recovery planning, \textbf{Action Hit} checks whether the predicted operation matches one valid operation, \textbf{Target Hit} checks whether the predicted target service matches a valid target, \textbf{Exact Match} requires both operation and target to match the gold plan, and \textbf{Recovery Validity} measures whether the predicted plan belongs to the incident-specific valid action set. \textbf{No-op Rate} is the fraction of predictions that return no action or an irrelevant action. 
Let $\mathcal{C}$ denote the event that both the root-cause service and fault type are correct, and let $\mathcal{V}$ denote the event that the predicted recovery plan is valid.
For post-diagnosis recovery analysis, we report conditional action validity $V_{\mathcal{C}}=\Pr(\mathcal{V}\mid\mathcal{C})$ and the invalid-action rate after correct diagnosis $E_{\mathcal{C}}=\Pr(\neg\mathcal{V}\mid\mathcal{C})=1-V_{\mathcal{C}}$.
$E_{\mathcal{C}}$ measures recovery-decision errors that remain after service and fault-type diagnosis are correct; lower values indicate stronger post-diagnosis recovery-action selection.
\section{Results}
\subsection{\researchq{1}: Diagnosis vs. Recovery-Action Validity}

\textbf{Experimental Setup.}
We first compare diagnostic correctness and recovery-action validity on the same incident population. Table~\ref{tab:diagnosis-recovery-validity} reports representative methods across two output interfaces. For recovery-oriented methods, the action metrics are computed from their own generated recovery plans. For RCA-only controls, the action metrics are computed through the fixed mapper described above and should be read as downstream recoverability under a transparent policy, not as the original methods' recovery-generation ability. 

\begin{table}[t]
\centering
\caption{Diagnosis-to-action evaluation for representative methods. Bold marks strong diagnosis, while red highlights low recovery-action validity.}
\label{tab:diagnosis-recovery-validity}
\small
\setlength{\tabcolsep}{3.5pt}
\resizebox{\linewidth}{!}{%
\begin{tabular}{lrrrrrrr}
\toprule
\textbf{Method} &
\textbf{RCA-S} $\uparrow$ &
\textbf{RCA-T} $\uparrow$ &
\textbf{Action} $\uparrow$ &
\textbf{Target} $\uparrow$ &
\textbf{Exact} $\uparrow$ &
\textbf{Valid} $\uparrow$ &
\textbf{No-op} $\downarrow$ \\
\midrule
\multicolumn{8}{l}{\textit{Recovery-oriented methods}} \\
LogPrompt & 0.603 & 0.162 & 0.228 & 0.513 & \textcolor{red!70!black}{0.000} & \textcolor{red!70!black}{0.109} & 0.470 \\
LogRAG & 0.616 & 0.142 & 0.228 & 0.546 & \textcolor{red!70!black}{0.000} & \textcolor{red!70!black}{0.123} & 0.447 \\
OpenRCA-D & 0.563 & 0.175 & 0.179 & 0.464 & \textcolor{red!70!black}{0.089} & \textcolor{red!70!black}{0.142} & 0.510 \\
Qwen-RAG & \textbf{0.990} & \textbf{0.957} & 0.719 & 0.709 & \textcolor{red!70!black}{0.258} & \textcolor{red!70!black}{0.483} & 0.119 \\
Kimi-RAG & \textbf{0.990} & \textbf{1.000} & 0.566 & 0.712 & \textcolor{red!70!black}{0.245} & \textcolor{red!70!black}{0.517} & 0.116 \\
MiniMax-RAG & \textbf{0.983} & \textbf{0.937} & 0.434 & 0.705 & \textcolor{red!70!black}{0.195} & \textcolor{red!70!black}{0.368} & 0.222 \\
DeepSeek-RAG & \textbf{0.914} & \textbf{0.934} & 0.838 & 0.666 & \textcolor{red!70!black}{0.222} & \textcolor{red!70!black}{0.530} & 0.103 \\
GLM-ZS & \textbf{0.997} & \textbf{0.993} & 0.321 & 0.712 & \textcolor{red!70!black}{0.073} & \textcolor{red!70!black}{0.315} & 0.288 \\
GLM-RAG & \textbf{0.997} & \textbf{0.997} & 0.722 & 0.709 & \textcolor{red!70!black}{0.242} & \textcolor{red!70!black}{0.603} & 0.043 \\
\midrule
\multicolumn{8}{l}{\textit{RCA-only + fixed mapper}} \\
Keyword & 0.116 & 0.063 & 0.149 & 0.103 & \textcolor{red!70!black}{0.026} & \textcolor{red!70!black}{0.089} & 0.775 \\
Rule & 0.338 & 0.228 & 0.298 & 0.278 & \textcolor{red!70!black}{0.139} & \textcolor{red!70!black}{0.222} & 0.642 \\
NB & \textbf{0.679} & \textbf{0.536} & 0.639 & 0.444 & \textcolor{red!70!black}{0.318} & \textcolor{red!70!black}{0.467} & 0.281 \\
Epsilon & 0.149 & 0.053 & 0.166 & 0.096 & \textcolor{red!70!black}{0.013} & \textcolor{red!70!black}{0.013} & 0.752 \\
RCD & 0.149 & 0.053 & 0.166 & 0.096 & \textcolor{red!70!black}{0.013} & \textcolor{red!70!black}{0.013} & 0.752 \\
Random Walk & 0.149 & 0.053 & 0.166 & 0.096 & \textcolor{red!70!black}{0.013} & \textcolor{red!70!black}{0.013} & 0.752 \\
BARO & 0.136 & 0.122 & 0.291 & 0.000 & \textcolor{red!70!black}{0.026} & \textcolor{red!70!black}{0.281} & 0.709 \\
CIRCA & 0.149 & 0.053 & 0.166 & 0.096 & \textcolor{red!70!black}{0.013} & \textcolor{red!70!black}{0.013} & 0.752 \\
LogFormer & 0.381 & 0.116 & 0.186 & 0.106 & \textcolor{red!70!black}{0.086} & \textcolor{red!70!black}{0.182} & 0.755 \\
OneLog & \textbf{0.533} & 0.086 & 0.169 & 0.166 & \textcolor{red!70!black}{0.109} & \textcolor{red!70!black}{0.152} & 0.732 \\
\bottomrule
\end{tabular}%
}
\end{table}

\begin{table*}[t]
  \centering
  \scriptsize
  \setlength{\tabcolsep}{2.6pt}
  \renewcommand{\arraystretch}{1.0}
  \caption{Multi-modal observation ablation. Each entry reports 5-fold mean $\pm$ standard deviation in percentage. Shaded cells mark the best input setting for each method and metric. Zero-shot and few-shot LLM backbones are omitted.}
  \label{tab:modality-ablation}

  \resizebox{\linewidth}{!}{
  \begin{tabular}{lcccccccccccc}
    \toprule
    \multirow{2}{*}{\textbf{Method}} &
    \multicolumn{4}{c}{\textbf{Logs}} &
    \multicolumn{4}{c}{\textbf{Logs+Events}} &
    \multicolumn{4}{c}{\textbf{Logs+Events+Metrics}} \\
    \cmidrule(lr){2-5}
    \cmidrule(lr){6-9}
    \cmidrule(lr){10-13}
    & \textbf{RCA-S} & \textbf{RCA-Type} & \textbf{Recovery} & \textbf{No-op}
    & \textbf{RCA-S} & \textbf{RCA-Type} & \textbf{Recovery} & \textbf{No-op}
    & \textbf{RCA-S} & \textbf{RCA-Type} & \textbf{Recovery} & \textbf{No-op} \\
    \midrule

    Rule
      & 3.3$\pm$1.2 & 10.3$\pm$0.8 & 12.9$\pm$1.3 & 72.5$\pm$5.2
      & \cellcolor{bestbg}\textbf{34.1$\pm$5.3} & 16.6$\pm$2.7 & \cellcolor{bestbg}\textbf{23.5$\pm$2.6} & 65.5$\pm$3.2
      & 33.8$\pm$5.5 & \cellcolor{bestbg}\textbf{22.9$\pm$3.6} & 22.2$\pm$2.4 & \cellcolor{bestbg}\textbf{64.2$\pm$4.4} \\

    NB
      & 17.9$\pm$3.0 & 16.2$\pm$5.0 & 6.6$\pm$4.3 & 61.9$\pm$8.3
      & 64.9$\pm$4.7 & 42.0$\pm$8.0 & 32.4$\pm$6.2 & 30.4$\pm$9.5
      & \cellcolor{bestbg}\textbf{68.6$\pm$3.3} & \cellcolor{bestbg}\textbf{52.7$\pm$6.0} & \cellcolor{bestbg}\textbf{46.0$\pm$5.2} & \cellcolor{bestbg}\textbf{28.8$\pm$6.2} \\

    LogRAG
      & 12.6$\pm$2.6 & 11.9$\pm$1.5 & 2.6$\pm$0.9 & 65.5$\pm$7.1
      & \cellcolor{bestbg}\textbf{59.6$\pm$6.6} & \cellcolor{bestbg}\textbf{13.9$\pm$3.7} & 10.3$\pm$4.6 & \cellcolor{bestbg}\textbf{46.0$\pm$2.9}
      & 59.3$\pm$7.8 & \cellcolor{bestbg}\textbf{13.9$\pm$3.3} & \cellcolor{bestbg}\textbf{11.9$\pm$4.0} & \cellcolor{bestbg}\textbf{46.0$\pm$4.2} \\

    LogPrompt
      & 15.6$\pm$1.0 & 11.6$\pm$2.1 & 2.3$\pm$2.2 & 68.9$\pm$3.6
      & 59.6$\pm$2.5 & 13.3$\pm$1.7 & 8.3$\pm$3.8 & 47.0$\pm$5.1
      & \cellcolor{bestbg}\textbf{61.6$\pm$3.0} & \cellcolor{bestbg}\textbf{16.3$\pm$3.7} & \cellcolor{bestbg}\textbf{12.3$\pm$2.6} & \cellcolor{bestbg}\textbf{46.0$\pm$4.2} \\

    OpenRCA-D
      & 12.6$\pm$2.8 & 10.9$\pm$1.5 & 4.3$\pm$1.9 & 75.5$\pm$2.9
      & 52.7$\pm$3.7 & 13.9$\pm$1.9 & 11.3$\pm$4.3 & 55.6$\pm$1.6
      & \cellcolor{bestbg}\textbf{55.6$\pm$2.7} & \cellcolor{bestbg}\textbf{15.2$\pm$1.5} & \cellcolor{bestbg}\textbf{12.2$\pm$3.0} & \cellcolor{bestbg}\textbf{52.6$\pm$3.0} \\

    Qwen-RAG
      & 93.1$\pm$3.1 & 93.4$\pm$1.7 & 32.1$\pm$4.8 & 24.8$\pm$2.8
      & \cellcolor{bestbg}\textbf{99.3$\pm$0.9} & \cellcolor{bestbg}\textbf{96.7$\pm$1.2} & 43.4$\pm$4.7 & 17.6$\pm$5.0
      & 99.0$\pm$0.9 & 96.4$\pm$1.4 & \cellcolor{bestbg}\textbf{48.0$\pm$3.2} & \cellcolor{bestbg}\textbf{11.9$\pm$3.0} \\

    Kimi-RAG
      & 97.4$\pm$1.5 & \cellcolor{bestbg}\textbf{100.0$\pm$0.0} & 54.6$\pm$3.7 & 9.9$\pm$2.6
      & \cellcolor{bestbg}\textbf{99.3$\pm$0.9} & \cellcolor{bestbg}\textbf{100.0$\pm$0.0} & \cellcolor{bestbg}\textbf{56.0$\pm$2.6} & \cellcolor{bestbg}\textbf{8.3$\pm$2.4}
      & 99.0$\pm$1.5 & \cellcolor{bestbg}\textbf{100.0$\pm$0.0} & 51.0$\pm$5.3 & 12.6$\pm$4.9 \\

    MiniMax-RAG
      & 71.8$\pm$42.5 & 69.8$\pm$41.3 & 23.9$\pm$14.3 & 43.5$\pm$33.1
      & \cellcolor{bestbg}\textbf{98.3$\pm$1.2} & \cellcolor{bestbg}\textbf{96.7$\pm$2.6} & 32.8$\pm$4.2 & 25.5$\pm$2.4
      & 97.7$\pm$1.9 & 92.4$\pm$2.5 & \cellcolor{bestbg}\textbf{36.1$\pm$2.8} & \cellcolor{bestbg}\textbf{22.9$\pm$2.9} \\

    DeepSeek-RAG
      & 85.7$\pm$7.3 & \cellcolor{bestbg}\textbf{97.3$\pm$1.9} & 54.5$\pm$11.5 & 8.3$\pm$5.1
      & \cellcolor{bestbg}\textbf{94.7$\pm$2.1} & 96.7$\pm$1.7 & 54.3$\pm$1.6 & 9.3$\pm$3.5
      & 94.1$\pm$3.6 & 96.7$\pm$2.3 & \cellcolor{bestbg}\textbf{54.7$\pm$5.4} & \cellcolor{bestbg}\textbf{7.0$\pm$2.1} \\

    GLM-RAG
      & 99.7$\pm$0.7 & \cellcolor{bestbg}\textbf{100.0$\pm$0.0} & \cellcolor{bestbg}\textbf{61.9$\pm$3.2} & \cellcolor{bestbg}\textbf{1.0$\pm$0.9}
      & \cellcolor{bestbg}\textbf{100.0$\pm$0.0} & \cellcolor{bestbg}\textbf{100.0$\pm$0.0} & 60.6$\pm$2.7 & 3.3$\pm$2.0
      & \cellcolor{bestbg}\textbf{100.0$\pm$0.0} & \cellcolor{bestbg}\textbf{100.0$\pm$0.0} & \cellcolor{bestbg}\textbf{61.9$\pm$3.4} & 2.0$\pm$0.7 \\

    \bottomrule
  \end{tabular}
  }
\end{table*}

\textbf{Results and Analysis.}
Table~\ref{tab:diagnosis-recovery-validity} shows that diagnostic metrics and recovery-action metrics measure different outputs. The strongest RAG-based LLMs achieve high RCA-Service accuracy, but their recovery metrics remain much lower. For example, Qwen-RAG reaches 0.990 RCA-Service and 0.957 RCA-Type, yet only 0.258 Exact Match and 0.483 Recovery Validity. Our results show that high-quality localization does not guarantee a valid operation-target decision. The action-level metrics also identify where recovery-action selection breaks. Some methods can often name a plausible target while failing to choose a valid operation or plan structure, as seen in LogPrompt, LogRAG, and OpenRCA-D. Other methods have low No-op rates but still limited Recovery Validity, showing that the problem is not merely refusal or empty output. Thus, Recovery Validity is not a redundant metric after RCA; it captures whether the method can translate evidence into an action that satisfies the incident-specific action space.

\rqresult{RQ1}{Diagnostic correctness and recovery-action validity diverge substantially on the same incident population. This divergence motivates action-level metrics for recovery-oriented methods, including operation hit, target hit, exact match, and recovery validity.}

\subsection{\researchq{2}: Modality Effects on Diagnosis and Recovery}

\textbf{Experimental Setup.}
To investigate how incident evidence affects diagnosis and recovery-action validity, we conduct a modality ablation study on methods that support the same three-view evaluation protocol. 
For each method, we evaluate three evidence settings: logs only, logs with Kubernetes events, and full evidence (logs, events, metrics, and runtime state). This allows us to assess how additional operational evidence affects diagnosis accuracy and recovery validity. Table~\ref{tab:modality-ablation} reports diagnosis accuracy, recovery validity, and no-op rate under each setting.

\textbf{Results and Analysis.}
Richer evidence generally improves recovery, but recovery-action validity remains an additional action-level metric. The effect is particularly evident for weaker methods. For example, NB improves from 6.6\% Recovery Validity with logs only to 46.0\% under full evidence, indicating that events, metrics, and runtime state provide substantial information beyond raw logs. Similar trends can also be observed for several LLM methods, although the magnitude of improvement varies across models.
Nevertheless, additional evidence alone does not determine action validity. For instance, with full evidence, LogPrompt achieves over 55\% RCA-Service accuracy, while its Recovery Validity remains close to 12\%, indicating that a correct diagnosis does not necessarily translate into executable recovery actions. Even among stronger RAG-constrained backbones, models can achieve near-perfect RCA-Service accuracy yet recover only about half of the incidents correctly.

\rqresult{RQ2}{Additional evidence improves both diagnosis and recovery in several methods, but recovery-action validity remains an additional action-level metric. The remaining errors indicate that action selection requires operation semantics and incident-specific constraints beyond richer observation alone.}

\subsection{\researchq{3}: Remaining Recovery Errors After Correct Diagnosis}

\textbf{Experimental Setup.}
To examine whether recovery failures mainly stem from diagnosis errors, we condition our analysis on correct RCA cases. We reconstruct the correct-RCA event $\mathcal{C}$ from service and fault-type labels, and measure $\Pr(\mathcal{C})$, $V_{\mathcal{C}}$, and $R_{\mathcal{C}}$. This analysis tests whether methods still fail to choose valid recovery operations and targets after correctly diagnosing the incident.

\begin{table}[t]
\centering
\caption{
Recovery errors after correct RCA. $\mathcal{C}$ denotes correct RCA, $V_{\mathcal{C}}=\Pr(\mathcal{V}\mid\mathcal{C})$, and $E_{\mathcal{C}}=1-V_{\mathcal{C}}$ is the invalid-action rate among correctly diagnosed incidents.
}
\label{tab:diagnosis-to-action-gap}
\scriptsize
\setlength{\tabcolsep}{1.8pt}
\resizebox{\linewidth}{!}{
\begin{tabular}{lrrrlrrr}
\toprule
\textbf{Method} & $\Pr(\mathcal{C})$ & $V_{\mathcal{C}}$ & $E_{\mathcal{C}}$ &
\textbf{Method} & $\Pr(\mathcal{C})$ & $V_{\mathcal{C}}$ & $E_{\mathcal{C}}$ \\
\midrule
Keyword & 0.000 & -- & -- & MiniMax-RAG & 0.924 & 0.380 & \textcolor{red!70!black}{0.620} \\
Rule-L & 0.010 & 1.000 & 0.000 & DeepSeek-ZS & 0.805 & 0.337 & \textcolor{red!70!black}{0.663} \\
Rule-LE & 0.126 & 1.000 & 0.000 & DeepSeek-FS & 0.728 & 0.409 & \textcolor{red!70!black}{0.591} \\
Rule-LEM & 0.212 & 0.984 & 0.016 & DeepSeek-RAG & 0.897 & 0.565 & \textcolor{red!70!black}{0.435} \\
\midrule
LogPrompt & 0.149 & 0.511 & 0.489 & GLM-ZS & 0.990 & 0.318 & 0.682 \\
LogRAG & 0.132 & 0.675 & 0.325 & GLM-FS & 0.944 & 0.354 & 0.646 \\
OpenRCA-D & 0.149 & 0.778 & 0.222 & GLM-RAG & \textbf{0.997} & 0.605 & \textcolor{red!70!black}{0.395} \\
OpenRCA-C & 0.000 & -- & -- & NB & 0.434 & 0.832 & 0.168 \\
\midrule
Qwen-ZS & 0.891 & 0.335 & \textcolor{red!70!black}{0.665} & Epsilon & 0.000 & -- & -- \\
Qwen-FS & 0.758 & 0.354 & \textcolor{red!70!black}{0.646} & Bayesian & 0.000 & -- & -- \\
Qwen-RAG & 0.947 & 0.500 & \textcolor{red!70!black}{0.500} & RCD & 0.000 & -- & -- \\
Kimi-ZS & 0.993 & 0.320 & \textcolor{red!70!black}{0.680} & RandomWalk & 0.000 & -- & -- \\
Kimi-FS & 0.907 & 0.380 & \textcolor{red!70!black}{0.620} & BARO & 0.000 & -- & -- \\
Kimi-RAG & 0.990 & 0.522 & \textcolor{red!70!black}{0.478} & CIRCA & 0.000 & -- & -- \\
MiniMax-ZS & 0.897 & 0.247 & \textcolor{red!70!black}{0.753} & LogFormer & 0.056 & 1.000 & 0.000 \\
MiniMax-FS & 0.825 & 0.285 & \textcolor{red!70!black}{0.715} & OneLog & 0.060 & 1.000 & 0.000 \\
\midrule
\multicolumn{4}{l}{\textit{Diagnostic control}} & Gold RCA + Mapper & 1.000 & 0.599 & \textcolor{red!70!black}{0.401} \\
\bottomrule
\end{tabular}}
\end{table}

\textbf{Results and Analysis.}
Table~\ref{tab:diagnosis-to-action-gap} shows substantial recovery errors even after correct RCA. The invalid-action rate gives direct evidence: when both the root-cause service and fault type are correctly identified, zero-shot and few-shot LLMs still leave $E_{\mathcal{C}}=0.591$--$0.753$, and RAG-based LLMs leave $E_{\mathcal{C}}=0.395$--$0.620$. These results isolate recovery failures from service- and fault-type misclassification. They show that identifying what failed provides only part of the information needed for recovery; the method must still select a valid operation and target.

RCA-only methods with Mapper serve as a diagnostic control. Their scores are determined by a fixed mapping policy, so they measure how far service and fault-type labels can be carried by a coarse service/type-to-action rule. Gold RCA + Mapper reaches $V_{\mathcal{C}}=0.599$, showing that service/type labels alone do not encode all fields required by an incident-specific operation-target plan. Valid recovery depends on finer-grained operational semantics, including configuration scopes, resource constraints, and operation preconditions.

\rqresult{RQ3}{Correct service and fault-type diagnosis still leaves many invalid recovery decisions. The invalid-action rate quantifies recovery-planning errors that are not explained by coarse RCA mistakes.}

\subsection{\researchq{4}: Error Sources in Recovery-Action Selection}

\textbf{Experimental Setup.}
To identify what makes recovery-action selection difficult, we analyze recovery failures from two complementary views. The first view categorizes invalid recovery predictions into operation, target, structure, and no-op errors. The second view groups predictions by fault type and compares RCA-Service accuracy with Recovery Validity. This analysis focuses on the five RAG-based LLM backbones because they achieve the strongest diagnostic performance, yet still exhibit a clear recovery-action gap.

\begin{table}[t]
\centering
\caption{recovery error taxonomy}
\label{tab:recovery-error-taxonomy}
\small
\resizebox{\linewidth}{!}{
\begin{tabular}{lrrp{3.5cm}}
\toprule
\textbf{Error Type} & \textbf{Count} & \textbf{Ratio} & \textbf{Interpretation} \\
\midrule
Wrong operation & 513 & 67.9\% & The method selects an operation that is not valid for the incident. \\
Invalid plan structure & 224 & 29.7\% & The operation is plausible, but the plan does not satisfy the recovery-plan semantics. \\
Wrong target/dependency & 10 & 1.3\% & The operation is correct, but the target service or dependency is invalid. \\
No-op or empty plan & 8 & 1.1\% & The method produces no executable recovery action. \\
\bottomrule
\end{tabular}}
\end{table}

\begin{figure}[t]
  \centering
  \includegraphics[width=0.8\linewidth]{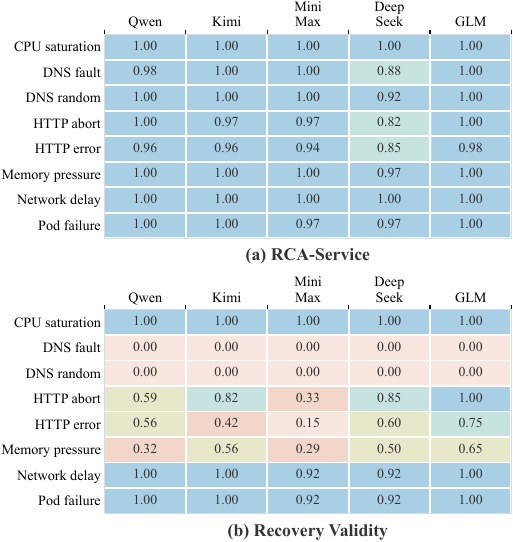}
  \caption{Per-fault difficulty across RAG-based LLM backbones. Panel (a) shows RCA-Service accuracy, while panel (b) shows recovery validity.}
  \label{fig:per-fault-difficulty}
\end{figure}

\textbf{Results and Analysis.}
Table~\ref{tab:recovery-error-taxonomy} identifies wrong operation selection as the dominant source of invalid plans, accounting for 67.9\% of all invalid predictions. Invalid plan structure contributes another 29.7\%, while no-op or empty plans and wrong target/dependency errors account for only 1.1\% and 1.3\%, respectively. The main difficulty therefore lies in choosing an incident-appropriate operation and expressing it in a valid recovery-plan structure, rather than merely deciding to act or locating the affected service.

Figure~\ref{fig:per-fault-difficulty} shows that this difficulty varies substantially across fault categories. Pod failure, CPU saturation, and network delay are recovered more reliably because their valid actions align with common operational routines such as restart or scale-out. DNS faults exhibit a stronger diagnosis-to-action gap: RCA-Service accuracy remains high across LLMs, ranging from 0.88 to 1.00, while recovery validity is 0.00 for both DNS categories. HTTP faults show similar semantic difficulty with larger model-dependent variation, where RCA-Service remains high but recovery validity ranges from 0.15 to 1.00 across models and subtypes. Memory pressure further exposes a resource-management failure mode, with strong service localization but recovery validity limited to 0.29--0.65.


\rqresult{RQ4}{Future recovery-oriented methods should complement service localization with explicit modeling of operation semantics, dependency scopes, configuration constraints, and resource preconditions.}

\subsection{\researchq{5}: Live Replay Check}

\textbf{Experimental Setup.}
To complement offline recovery validity with live execution behavior, we replay model-generated recovery plans on a live Kubernetes cluster. We use Qwen-RAG as the representative method and replay the full 302-incident benchmark. For each case, we inject the original Chaos Mesh fault into the cluster, execute the predicted recovery action, and check post-recovery system health within a 75-second window. To keep the online check aligned with the offline metric, Table~\ref{tab:online-replay-outcomes} reports a validity-gated replay outcome: a case is counted as recovered only when the prediction satisfies the offline recovery-validity specification and the replay restores service health. 

\textbf{Results and Analysis.}
Table~\ref{tab:online-replay-outcomes} shows that validity-gated live replay is consistent with offline Recovery Validity. Among the 302 replayed incidents, 146 Qwen-RAG predictions are both offline-valid and successful in live execution, giving a replay-consistent recovery rate of 48.3\%. Since replay re-injects the original Chaos Mesh fault and executes the predicted operation in the cluster, this result confirms that offline-valid plans are executable and can restore service health under the replay protocol. The per-fault results follow the offline error pattern. CPU saturation, network delay, and pod failure reach 100.0\% replay RV because their repairs correspond to direct routines such as scale-out or restart. In contrast, DNS faults remain at 0.0\% because the predicted plans often omit the required dependency target. HTTP and memory-pressure cases recover only partially, reflecting the need for configuration- or resource-level recovery semantics. Overall, live replay supports the main finding that the bottleneck is producing a complete operation-target plan, rather than merely executing an already valid action.

\begin{table}[t]
  \centering
  \caption{Validity-gated live replay outcomes for Qwen-RAG over the full 302-incident benchmark.}
  \label{tab:online-replay-outcomes}
  \scriptsize
  \resizebox{\linewidth}{!}{
  \begin{tabular}{lrrrr}
    \toprule
    Fault type & Plan & RV-valid & Rec. & Replay RV \\
    \midrule
    CPU saturation   & 36 & 36 & 36 & 100.0\% \\
    DNS fault        & 48 & 0  & 0  & \textcolor{red!70!black}{0.0\%} \\
    DNS random       & 48 & 0  & 0  & \textcolor{red!70!black}{0.0\%} \\
    HTTP abort       & 39 & 23 & 23 & \textcolor{red!70!black}{59.0\%} \\
    HTTP error       & 48 & 27 & 27 & \textcolor{red!70!black}{56.2\%} \\
    Memory pressure  & 34 & 11 & 11 & \textcolor{red!70!black}{32.4\%} \\
    Network delay    & 13 & 13 & 13 & 100.0\% \\
    Pod failure      & 36 & 36 & 36 & 100.0\% \\
    \midrule
    \textbf{Total}   & \textbf{302} & \textbf{146} & \textbf{146} & \textbf{48.3\%} \\
    \bottomrule
  \end{tabular}}
  \vspace{0.2em}
  \footnotesize
  \emph{Plan}: planned attempts; \emph{RV-valid}: predictions satisfying the offline recovery-validity specification; \emph{Rec.}: offline-valid predictions that restore service health in replay; \emph{Replay RV}: validity-gated replay recovery rate.
\end{table}

\rqresult{RQ5}{Validity-gated live replay confirms the offline recovery-validity signal, indicating that invalid recovery mainly comes from incomplete operation-target planning rather than execution failure.}

\textbf{Supplementary Material.}
More details and results are available in the supplementary material, including the full dataset, annotation schema, recovery-plan semantics, method implementation details, and additional analyses.

\section{Discussion}

\textbf{Recovery Requires Additional Action-Level Metrics.} R2Act does not argue that RCA benchmarks should evaluate recovery, nor that RCA methods are expected to solve recovery by themselves. Diagnosis and recovery can use the same incidents and evidence, but they answer different questions: RCA metrics ask what failed, while recovery-action metrics ask whether the selected operation and target can repair the incident. RQ1 and RQ3 show that this distinction matters because recovery-action validity diverges from diagnostic correctness even after correct RCA.

\textbf{Recovery Depends on Action Semantics.} The per-fault results show why diagnosis alone is insufficient. Lifecycle and some network cases align with direct routines such as restart or scale-out, but DNS, HTTP, and memory cases require reasoning about dependencies, configuration scopes, or resource constraints. These cases show that recovery actions should be modeled as typed operational decisions, not as free-form recommendations attached to RCA outputs.

\textbf{More Evidence Helps but Is Insufficient.} RQ2 shows that events, metrics, and runtime state improve recovery validity for several methods. Yet Recovery Validity remains far below RCA-Service even with full evidence, indicating that many failures come from mapping evidence to a valid operation and target rather than from missing observations alone.


\textbf{Offline Validity Needs Execution Checks.} The online replay experiment connects offline recovery validity with execution over the full 302-incident benchmark: 146 Qwen-RAG predictions are both offline-valid and recovered in live execution. This does not collapse the diagnosis-to-action gap. Offline RV remains the controlling metric because it checks whether the predicted plan contains the required operation, target, and incident-specific fields before replay success is counted. The two checks are therefore complementary: offline labels encode operational constraints, and replay verifies restored health under those constraints.

\textbf{Scope and Limitations.} The current benchmark instance is built on \system and covers six service roles and eight fault categories. As a controlled testbed, it does not yet capture multiple production systems, larger service graphs, or organization-specific recovery policies. To support extension, R2Act separates the benchmark schema, quality gates, and evaluator from the particular system used in the current release. New systems, fault campaigns, and recovery actions can therefore be appended through the same pipeline. The current release prioritizes auditable recovery under controlled fault injection, while the schema and evaluator also support expanding with additional operator-validated equivalent plans.


\section{Conclusion}
This paper introduces \benchmark, a recovery-action evaluation framework for microservice failure recovery. We instantiate it as a 302-incident benchmark dataset that connects multi-modal incident evidence, RCA labels, incident-specific action spaces, recovery-plan annotations, and selected live replay evidence. RCA-centered metrics remain appropriate for diagnosis, but they do not determine whether a method can choose an admissible operation-target plan. The observed failures are driven mainly by operation-selection and plan-structure errors in faults that require dependency, configuration, or resource semantics. These findings support judging future recovery-oriented methods not only by diagnostic correctness, but also by whether their actions are valid, executable, and able to restore post-recovery health.
Full live replay further shows that offline action validity needs execution checks because cluster readiness, executor coverage, and health-check granularity can affect whether an executed plan restores service health.
Together, these results position recovery validity as a first-class objective for recovery-oriented incident-response evaluation.

\bibliographystyle{IEEEtran}
\bibliography{references}

\end{document}